\RequirePackage{lineno}
\documentclass[twocolumn,showpacs,preprintnumbers,amsmath,amssymb,prl,superscriptaddress]{revtex4}
\usepackage{graphicx}
\usepackage{dcolumn}
\usepackage{bm}
\modulolinenumbers[1]
\newenvironment{color}[3]
{
}{
}
\newcommand{\grey}[1]     {}

\begin{document}
\title{Energy Dependence of Moments of Net-proton Multiplicity Distributions at RHIC}
\affiliation{AGH University of Science and Technology, Cracow, Poland}
\affiliation{Argonne National Laboratory, Argonne, Illinois 60439, USA}
\affiliation{University of Birmingham, Birmingham, United Kingdom}
\affiliation{Brookhaven National Laboratory, Upton, New York 11973, USA}
\affiliation{University of California, Berkeley, California 94720, USA}
\affiliation{University of California, Davis, California 95616, USA}
\affiliation{University of California, Los Angeles, California 90095, USA}
\affiliation{Universidade Estadual de Campinas, Sao Paulo, Brazil}
\affiliation{Central China Normal University (HZNU), Wuhan 430079, China}
\affiliation{University of Illinois at Chicago, Chicago, Illinois 60607, USA}
\affiliation{Cracow University of Technology, Cracow, Poland}
\affiliation{Creighton University, Omaha, Nebraska 68178, USA}
\affiliation{Czech Technical University in Prague, FNSPE, Prague, 115 19, Czech Republic}
\affiliation{Nuclear Physics Institute AS CR, 250 68 \v{R}e\v{z}/Prague, Czech Republic}
\affiliation{Frankfurt Institute for Advanced Studies FIAS, Germany}
\affiliation{Institute of Physics, Bhubaneswar 751005, India}
\affiliation{Indian Institute of Technology, Mumbai, India}
\affiliation{Indiana University, Bloomington, Indiana 47408, USA}
\affiliation{Alikhanov Institute for Theoretical and Experimental Physics, Moscow, Russia}
\affiliation{University of Jammu, Jammu 180001, India}
\affiliation{Joint Institute for Nuclear Research, Dubna, 141 980, Russia}
\affiliation{Kent State University, Kent, Ohio 44242, USA}
\affiliation{University of Kentucky, Lexington, Kentucky, 40506-0055, USA}
\affiliation{Korea Institute of Science and Technology Information, Daejeon, Korea}
\affiliation{Institute of Modern Physics, Lanzhou, China}
\affiliation{Lawrence Berkeley National Laboratory, Berkeley, California 94720, USA}
\affiliation{Massachusetts Institute of Technology, Cambridge, MA 02139-4307, USA}
\affiliation{Max-Planck-Institut f\"ur Physik, Munich, Germany}
\affiliation{Michigan State University, East Lansing, Michigan 48824, USA}
\affiliation{Moscow Engineering Physics Institute, Moscow Russia}
\affiliation{National Institute of Science Education and Research, Bhubaneswar 751005, India}
\affiliation{Ohio State University, Columbus, Ohio 43210, USA}
\affiliation{Old Dominion University, Norfolk, VA, 23529, USA}
\affiliation{Institute of Nuclear Physics PAN, Cracow, Poland}
\affiliation{Panjab University, Chandigarh 160014, India}
\affiliation{Pennsylvania State University, University Park, Pennsylvania 16802, USA}
\affiliation{Institute of High Energy Physics, Protvino, Russia}
\affiliation{Purdue University, West Lafayette, Indiana 47907, USA}
\affiliation{Pusan National University, Pusan, Republic of Korea}
\affiliation{University of Rajasthan, Jaipur 302004, India}
\affiliation{Rice University, Houston, Texas 77251, USA}
\affiliation{Universidade de Sao Paulo, Sao Paulo, Brazil}
\affiliation{University of Science \& Technology of China, Hefei 230026, China}
\affiliation{Shandong University, Jinan, Shandong 250100, China}
\affiliation{Shanghai Institute of Applied Physics, Shanghai 201800, China}
\affiliation{SUBATECH, Nantes, France}
\affiliation{Temple University, Philadelphia, Pennsylvania, 19122, USA}
\affiliation{Texas A\&M University, College Station, Texas 77843, USA}
\affiliation{University of Texas, Austin, Texas 78712, USA}
\affiliation{University of Houston, Houston, TX, 77204, USA}
\affiliation{Tsinghua University, Beijing 100084, China}
\affiliation{United States Naval Academy, Annapolis, MD 21402, USA}
\affiliation{Valparaiso University, Valparaiso, Indiana 46383, USA}
\affiliation{Variable Energy Cyclotron Centre, Kolkata 700064, India}
\affiliation{Warsaw University of Technology, Warsaw, Poland}
\affiliation{University of Washington, Seattle, Washington 98195, USA}
\affiliation{Yale University, New Haven, Connecticut 06520, USA}
\affiliation{University of Zagreb, Zagreb, HR-10002, Croatia}

\author{L.~Adamczyk}\affiliation{AGH University of Science and Technology, Cracow, Poland}
\author{J.~K.~Adkins}\affiliation{University of Kentucky, Lexington, Kentucky, 40506-0055, USA}
\author{G.~Agakishiev}\affiliation{Joint Institute for Nuclear Research, Dubna, 141 980, Russia}
\author{M.~M.~Aggarwal}\affiliation{Panjab University, Chandigarh 160014, India}
\author{Z.~Ahammed}\affiliation{Variable Energy Cyclotron Centre, Kolkata 700064, India}
\author{I.~Alekseev}\affiliation{Alikhanov Institute for Theoretical and Experimental Physics, Moscow, Russia}
\author{J.~Alford}\affiliation{Kent State University, Kent, Ohio 44242, USA}
\author{C.~D.~Anson}\affiliation{Ohio State University, Columbus, Ohio 43210, USA}
\author{A.~Aparin}\affiliation{Joint Institute for Nuclear Research, Dubna, 141 980, Russia}
\author{D.~Arkhipkin}\affiliation{Brookhaven National Laboratory, Upton, New York 11973, USA}
% ...

\medskip
\collaboration{STAR Collaboration}\noaffiliation

\date{\today}
\begin{abstract}
We report the beam energy ($\sqrt{s_{\mathrm {NN}}}$ = 7.7 - 200 GeV) and collision centrality 
dependence of the mean ($M$), standard deviation ($\sigma$), skewness ($S$), and kurtosis ($\kappa$) 
of the net-proton multiplicity distributions in Au+Au collisions. The measurements are carried out 
by the STAR experiment at midrapidity ($|y| < 0.5$) and within the transverse momentum range 
0.4 $<$ $p_{\rm T}$ $<$ 0.8 GeV/$c$ in the first phase of the Beam Energy Scan program at the Relativistic 
Heavy Ion Collider. These measurements are important for understanding the Quantum Chromodynamic (QCD) phase diagram. The products 
of the moments, $S\sigma$ and $\kappa\sigma^{2}$, are sensitive to the correlation length of the hot 
and dense medium created in the collisions and are related to the ratios of baryon number susceptibilities 
of corresponding orders. The products of moments are found to have values significantly below the 
Skellam expectation and close to expectations based on independent proton and anti-proton production.
The measurements are compared to a transport model calculation to understand the effect of acceptance 
and baryon number conservation, and also to a hadron resonance gas model. 
\end{abstract}
\maketitle
The Beam Energy Scan (BES) program at the Relativistic Heavy-Ion Collider (RHIC)
facility aims at studying in detail the QCD phase structure. This enables
us to map the phase diagram, temperature ($T$) versus baryonic chemical potential 
($\mu_{\rm B}$), of strong interactions. Important advancements have been made towards 
the understanding of the QCD phase structure at small $\mu_{\rm B}$. Theoretically, it 
has been found that at high temperatures, there occurs a cross-over transition from 
hadronic matter to a de-confined state of quarks and gluons at  $\mu_{\rm B}$ = 0 MeV~\cite{Aoki:2006we}. 
Experimental data from RHIC and the Large Hadron Collider have provided evidence of 
the formation of QCD matter with quark and gluon degrees of freedom~\cite{starwhitepaper}. 
Several studies have been done to estimate the quark-hadron transition temperature 
at $\mu_{\rm B}$ = 0~\cite{transitiontemp}. Interesting features of the QCD phase structure 
are expected to appear at larger $\mu_{\rm B}$~\cite{Fukushima:2010bq}. These include the QCD critical point (CP)
~\cite{qcp,qcp1} and a first order phase boundary between quark-gluon and hadronic phases~\cite{firstorder}. 

Previous studies of net-proton multiplicity 
distributions suggest that the possible CP region is unlikely to be below $\mu_{\rm B}$ = 200 MeV~\cite{starkurtosisprl}.
The versatility of the RHIC machine has permitted the center of mass energy ($\sqrt{s_{\mathrm {NN}}}$) to
be varied below the injection energy ($\sqrt{s_{\mathrm {NN}}}$ = 19.6 GeV), thereby providing the 
possibility to scan the QCD phase diagram above $\mu_{\rm B}$ $\sim$ 250 MeV.
The  $\mu_{\rm B}$ value is observed to increase with decreasing $\sqrt{s_{\mathrm {NN}}}$~\cite{cleymans}.
The goal of the BES program at RHIC is to look for the experimental signatures of a first order phase 
transition and the critical point by colliding Au ions at various $\sqrt{s_{\mathrm {NN}}}$~\cite{bes}.

Non-monotonic variations of observables related to the moments of the 
distributions of conserved quantities such as net-baryon, net-charge, 
and net-strangeness~\cite{volker} number with $\sqrt{s_{\mathrm {NN}}}$ are
believed to be good signatures of a phase transition and a CP. 
The moments are related to the correlation length ($\xi$) of the system~\cite{stephanovmom}.
The signatures of phase transition or CP are detectable if they 
survive the evolution of the system~\cite{survival}. Finite size and 
time effects in heavy-ion collisions put constraints 
on the significance of the desired signals. A theoretical calculation suggests 
a non-equilibrium $\xi$ $\approx$ 2-3 fm for heavy-ion collisions~\cite{krishnaxi}. Hence, it is
proposed to study the higher moments (like skewness, 
${\it {S}}$ = $\left\langle (\delta N)^3 \right\rangle/\sigma^{3}$ 
and kurtosis, $\kappa$ = [$\left\langle (\delta N)^4 \right\rangle/\sigma^{4}$] -- 3 
with $\delta N$ = $N$ -- $\langle N \rangle$) of distributions of conserved quantities
due to a stronger dependence on $\xi$~\cite{stephanovmom}. Both the magnitude
and the sign of the moments~\cite{asakawa}, which quantify the shape of the multiplicity
distributions, are important for understanding phase transition and CP effects.
Further, products of the moments can be related to susceptibilities associated 
with the conserved numbers. The product $\kappa$$\sigma^2$ of the net-baryon number
distribution is related to the ratio of fourth order ($\chi^{(4)}_{\mathrm B}$) 
to second order ($\chi^{(2)}_{\mathrm B}$) baryon number susceptibilities~\cite{latticesus,Gavai:2010zn}. 
The ratio $\chi^{(4)}_{\mathrm B}$/$\chi^{(2)}_{\mathrm B}$ is expected to deviate
from unity near the CP. It has different values for the hadronic and partonic phases~\cite{Gavai:2010zn}. 

This Letter reports measurements of the energy dependence of 
higher moments of the net-proton multiplicity
($N_{p} - N_{\bar{p}}$ = $\Delta N_{p}$) 
distributions from Au+Au collisions. The aim is to search for signatures of the CP 
over a broad range of $\mu_{B}$ in the QCD phase diagram. Theoretical calculations 
have shown that $\Delta N_{p}$ fluctuations reflect
the singularity of the charge and baryon number susceptibility, as expected at
the CP~\cite{hatta}. The measurements presented here are within a finite acceptance 
range and only use the protons among the produced baryons. 
Refs.~\cite{Kitazawa:2012at,Bzdak:2012ab} discuss the advantages of using net-baryon
measurements and effects of acceptance on which the measurements depend intrinsically (e.g. conservation laws 
and other finite statistical fluctuations dominate near full and small acceptance respectively).

\begin{figure}[htp]
\includegraphics[scale=0.43]{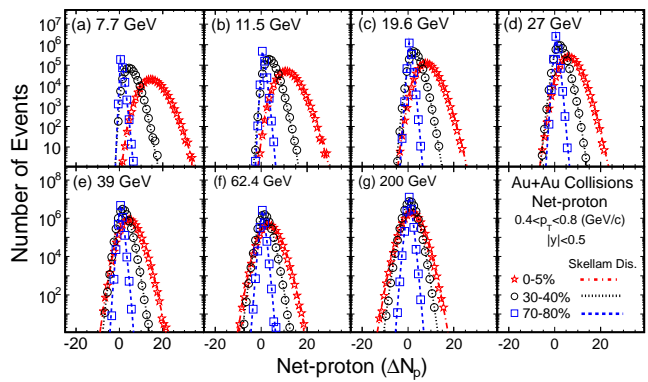}
\caption
{(Color online) 
$\Delta N_{p}$ multiplicity distributions in Au+Au collisions at 
various $\sqrt{s_{\mathrm {NN}}}$ for 0-5\%, 30-40\% and 70-80\% collision centralities 
at midrapidity. The statistical errors are small and within the symbol size. 
The lines are the corresponding Skellam distributions. 
The distributions are not corrected for the finite centrality width effect and $N_{p}$($N_{\bar p}$) 
reconstruction  efficiency. 
}
\label{Fig1}
\end{figure}
The data presented in the paper were obtained using the Time Projection
Chamber (TPC) of the Solenoidal Tracker at RHIC (STAR)~\cite{star}. 
The event-by-event proton ($N_{p}$) and anti-proton ($N_{\bar{p}}$) multiplicities
are measured for Au+Au minimum-bias events at $\sqrt{s_{\mathrm {NN}}}$ = 11.5, 19.6, 27, 39, 
62.4, and 200 GeV for collisions occurring within $\Delta Z$ = 30 cm from 
the TPC center along the beam line. For 7.7 GeV, $\Delta Z$ is 50 cm.  
The 19.6 and 27 GeV data were collected in the year 2011 and the other energies were taken in 2010.
Interactions of the beam with the beam pipe are rejected by choosing events
with a radial vertex position in the transverse plane of less than 2 cm.
The numbers of events analyzed are 3$\times 10^{6}$, 6.6$\times 10^{6}$, 15$\times 10^{6}$, 30$\times 10^{6}$, 
86$\times 10^{6}$, 47$\times 10^{6}$, and 238$\times 10^{6}$ for  
$\sqrt{s_{\mathrm {NN}}}$ = 7.7, 11.5, 19.6, 27, 39, 62.4, and 200 GeV, respectively. 
Similar studies have also been carried out in $p$+$p$ collisions with 
0.6 $\times 10^{6}$ and 7$\times 10^{6}$ events at $\sqrt{s_{\mathrm {NN}}}$ = 62.4 and 200 GeV, respectively. 
The centrality selection  utilizes the uncorrected charged particle multiplicity other than identified protons 
and anti-protons within pseudorapidity $|\eta|$ $<$ 1.0 measured by the TPC. It is found that the measured 
net-proton moment values depend on the choice of the pseudorapidity range for the centrality selection. 
However the values of the moments do not change if the centrality selection range is further increased to the full
acceptance of the TPC (which leads to a 15\% increase in charged particle multiplicity).
In the UrQMD~\cite{urqmd} studies, after increasing the $\eta$ range used for 
centrality selection to two units, it is observed that the maximum decrease of moments is $\sim$ 2.5\% and 35\% 
for $\sqrt{s_{\mathrm {NN}}}$ = 200 and 7.7 GeV, respectively~\cite{Luo:2013bmi}. There is minimal change
for central collisions compared to other centralities.
For each centrality, the average number of participants ($\langle N_{\mathrm {part}} \rangle$) is
obtained by Glauber model calculations. The $\Delta N_{p}$ measurements are 
carried out at midrapidity ($|y|$ $<$ 0.5) in the range 
0.4 $<$ $p_{\mathrm T}$ $<$ 0.8 GeV/$c$. Ionization energy loss ($dE/dx$) of charged 
particles in the TPC is used to identify the inclusive $p$($\bar{p}$)~\cite{starprc}. 
The minimum $p_{\rm T}$ cut and a maximum distance of closest approach (DCA)
to the collision vertex of 1 cm for each $p(\bar{p})$
candidate track suppress contamination from secondaries~\cite{starprc}.
To have a good purity of the proton sample (better than 98\%) for all beam energies, the maximum
$p_{\mathrm T}$ is taken to be 0.8 GeV/$c$.
This $p_{\mathrm T}$ interval accounts for 
approximately 50\% of the total uncorrected $p+ \bar{p}$ multiplicity at midrapidity. 
The average proton reconstruction efficiency for the $p_{\mathrm T}$ range studied is between 70-78\% 
and 83-86\%, for central and peripheral collisions, respectively, at different $\sqrt{s_{\mathrm {NN}}}$.

$\Delta N_{p}$ distributions from 70-80\%, 30-40\%, and 0-5\% 
Au+Au collision centralities are shown in Fig.~\ref{Fig1}. 
The $\Delta N_{p}$ is not corrected for reconstruction efficiency. 
The distributions are also not corrected for the finite centrality
width effect~\cite{Luo:2013bmi}.  The subsequent analysis in this Letter 
is corrected for the centrality width effect.
The corresponding Skellam distributions are also shown, $P(N) = \left( \frac{M_{p}}{M_{\bar{p}}}\right)^{N/ 2}{{I_N(2\sqrt {M_{p}M_{\bar{p}}}~)} ~{\exp[-(M_{p} +{M_{\bar p}})]}} \;$,
where $I_N(x)$ is a modified Bessel function of the first kind, and $M_{p}$ and $M_{\bar p}$
are the measured mean multiplicities of proton and anti-protons~\cite{BraunMunzinger:2011dn}. 
The data seems to closely follow the Skellam distributions. To study the detail shape of the distribution, 
we discuss the various order cumulants ($C_{n}$), where $C_{1}$ = $M$, $C_{2}$ = $\sigma^2$,
$C_{3}$ = $S\sigma^{3}$ and $C_{4}$ = $\kappa\sigma^{4}$. 
For both proton and anti-proton distributions being Poissonian, the $\Delta N_{\mathrm p}$ distribution 
will be a Skellam and have $C_{3}/C_{2}$ = $(M_{p} - M_{\bar p})/(M_{p} + M_{\bar p})$ and  $C_{4}/C_{2}$ = 1.

\begin{figure}[htp]
\includegraphics[width=0.46\textwidth]{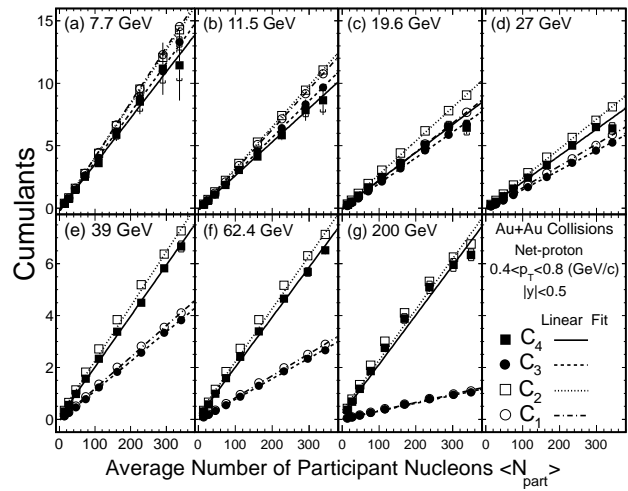}
\caption
{Centrality dependence of the cumulants of $\Delta N_{p}$ distributions
for Au+Au collisions. Error bars are statistical and caps are systematic errors.
}
\label{Fig2}
\end{figure}

\begin{figure*}[htp]
\includegraphics[scale=0.6]{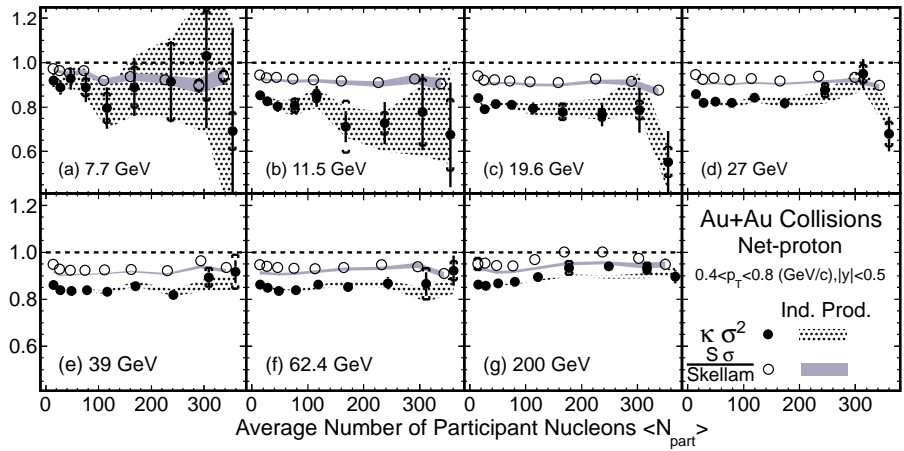}
\caption
{(Color online) Centrality dependence of ${\it{S}}$$\sigma$/Skellam
and $\kappa$$\sigma^2$ for $\Delta N_{p}$ in Au+Au collisions at 
$\sqrt{s_{\mathrm {NN}}}$ = 7.7, 11.5, 19.6, 27, 39, 62.4, and 200 GeV. 
The results are corrected for the $p(\bar{p})$ reconstruction efficiency.
The error bars are statistical and caps are systematic errors. 
The shaded bands are expectations assuming the approach of
independent proton and anti-proton production, as described in the text.  
The width of the bands represents statistical uncertainty.}
\label{Fig3}
\end{figure*}
The four cumulants that describe the shape of $\Delta N_{p}$ distributions at various collision
energies are plotted as a function of $\langle N_{\mathrm {part}} \rangle$ in Fig.~\ref{Fig2}. 
We use the Delta theorem approach to obtain statistical errors~\cite{Luo:2011tp}.
The typical statistical error values for $C_{2}$, $C_{3}$, and $C_{4}$ for central Au+Au collisions at 7.7 GeV 
are 0.3\%, 2.5\% and 2.5\%
respectively, and  those for high statistics 200 GeV results are 0.04\%, 1.2\% and 2.0\% respectively.
Most of the cumulant values show a linear variation with $\langle N_{\mathrm {part}}\rangle$.
The $C_{1}$ values increase as $\sqrt{s_{\mathrm {NN}}}$ decreases, in accordance with the energy
and centrality dependence of baryon transport. 
$C_{2}$ and $C_{4}$ have similar values as a function of  $\langle N_{\mathrm {part}}\rangle$ for a given $\sqrt{s_{\mathrm {NN}}}$.
$C_{1}$ and $C_{3}$ follow each other closely as a function of $\langle N_{\mathrm {part}}\rangle$ 
at any given $\sqrt{s_{\mathrm {NN}}}$. The differences between these groupings decrease as $\sqrt{s_{\mathrm {NN}}}$ decreases. 
The decrease in the $C_{3}$ values with increasing beam energy indicates that the distributions become symmetric for 
the higher beam energies. 
The particle production at any given centrality can be considered a superposition of 
several identically distributed independent sources the number of which is proportional to $N_{\mathrm {part}}$~\cite{starkurtosisprl}. 
For the cumulants, this means
a linear increase with $\langle N_{\mathrm {part}}\rangle$ as the system volume increases. This reflects that the cumulants are
extensive quantities that are proportional to system volume. The lines in  
Fig.~\ref{Fig2} are linear fits to the cumulants, which provide a reasonable description of the centrality dependence.
This indicates that the volume effect dominates the measured cumulants values. The $\chi^{2}/\rm{ndf}$ between the 
linear fit and data are smaller than 3.2 for all cumulants presented. The slight deviation of some cumulants in most central collisions 
from the fit line are due to the corresponding proton distributions.

In order to cancel the volume effect to first order and to understand the collision dynamics,  we present the ratios 
of the cumulants $C_{3}/C_{2}$ (= ${\it{S}}$$\sigma$) and $C_{4}/C_{2}$ (= $\kappa$$\sigma^2$) as a function of $\langle N_{\mathrm {part}} \rangle$ 
for all collision energies, in Fig.~\ref{Fig3}. The ${\it{S}}$$\sigma$ are normalized to the
corresponding Skellam expectations. Results with correction for the $p(\bar{p})$ reconstruction efficiency
are presented.  The correction for a finite track reconstruction efficiency is done by assuming a binomial distribution for 
the probability to reconstruct $n$ particles out of $N$ produced~\cite{eff,Bzdak:2012ab}.
These observables are related to 
the ratio of baryon number susceptibilities ($\chi_{\mathrm B}$) at a given temperature ($T$) 
computed in QCD motivated models as: ${\it{S}}$$\sigma$ = $(\chi^{(3)}_{\mathrm B}/T)/(\chi^{(2)}_{\mathrm B}/T^2)$ and
$\kappa$$\sigma^2$ = $(\chi^{(4)}_{\mathrm B})/(\chi^{(2)}_{\mathrm B}/T^2)$~\cite{latticesus,Gavai:2010zn}. 
Close to the CP, QCD based calculations predict the net-baryon number distributions to be non-Gaussian and 
susceptibilities to diverge, causing ${\it{S}}$$\sigma$ and $\kappa$$\sigma^2$ to have non-monotonic variations
with  $\langle N_{\mathrm {part}}\rangle$ and/or  $\sqrt{s_{\mathrm {NN}}}$~\cite{qcp1,stephanovmom}.

We observe in Fig.~\ref{Fig3} the $\kappa$$\sigma^2$  and the ${\it{S}}$$\sigma$  normalized to Skellam expectations 
are below unity for all of the Au+Au collision data sets presented. 
The deviations below unity of the order of 1-3\%~\cite{stephanovprd} as seen for the central collisions for energies
above 27 GeV are expected from quantum statistical effects. 
The measured ${\it{S}}$$\sigma$ and $\kappa$$\sigma^2$ are compared to expectations 
in which the cumulants of $\Delta N_{p}$ distributions are constructed by considering independent  
production of protons and anti-protons. For independent production,
the various order ($n$ = 1, 2, 3 and 4) net-proton cumulants are given as $C_{n}(\Delta N_{p})$ = $C_{n}(N_{p})$ + $(-1)^{n}$$C_{n}(N_{{\bar{p}}})$, where $C_{n}(N_{p})$ and $C_{n}(N_{{\bar{p}}})$ are cumulants of the measured distributions 
of $N_{\mathrm p}$  and $N_{{\bar{p}}}$, respectively. This approach breaks intra-event correlations between $N_{p}$ and $N_{{\bar {p}}}$.
The results from independent production are found to be in good agreement with the data. 
However, for $\sqrt{s_{\mathrm {NN}}}$ $<$ 39 GeV, the $C_{n}$ of net-protons are dominated  by the corresponding values from the proton distributions.
The assumption that $N_p$ and $N_{\bar{p}}$ have independent binomial distributions~\cite{Tarnowsky:2012vu} 
also leads to a good description of the measurements (similar to independent production, but not plotted in Fig.~\ref{Fig3}). 

Systematic errors are estimated by varying the following requirements 
for $p(\bar{p})$ tracks: DCA, track quality reflected by the number of fit points used in track reconstruction,
and the $dE/dx$ selection criteria for $p(\bar{p})$ 
identification. The typical systematic errors are of the order 4\% for $\it {M}$ and $\sigma$, 5\% for ${\it{S}}$  
and 12\% for $\kappa$. A 5\% uncertainty in reconstruction efficiency estimation is also considered. 
The statistical and systematic (caps) errors are presented separately in the figures.

\begin{figure}[htp]
\includegraphics[width=0.45\textwidth]{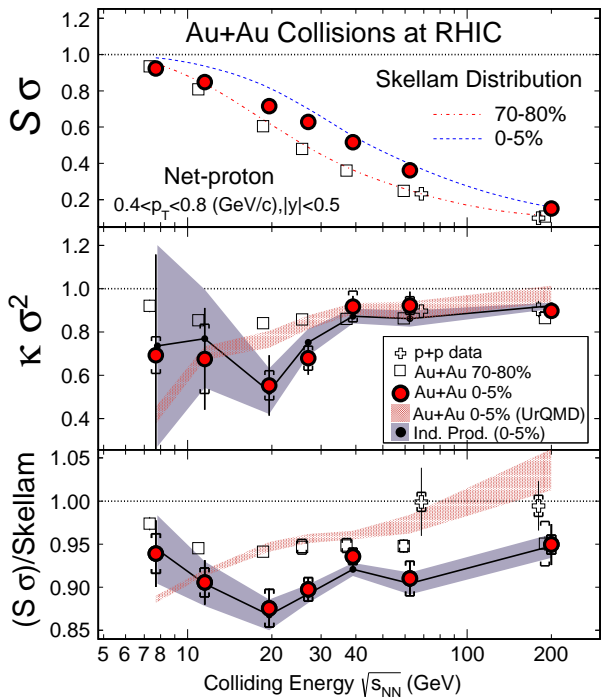}
\caption
{(Color online) Collision energy and centrality dependence of the net-proton ${\it{S}}\sigma$ and $\kappa$$\sigma^2$
from Au+Au and $p$+$p$ collisions at RHIC. 
Crosses, open squares and filled circles are for the efficiency corrected results of $p$+$p$, 70-80\%,
and 0-5\% Au+Au collisions, respectively. Skellam distributions for corresponding 
collision centralities are shown in the top panel. Shaded hatched bands are the results from
UrQMD~\cite{urqmd}. In the middle and lower panels, the shaded solid bands are 
the expectations assuming independent proton and anti-proton production. 
The width of the bands represents statistical uncertainties. 
The hadron resonance gas model (HRG) values for $\kappa$$\sigma^2$ and ${\it{S}}\sigma$/Skellam are unity. 
The error bars are statistical and caps are systematic errors. 
For clarity, $p$+$p$ and 70-80\% Au+Au results are slightly displaced horizontally.
}
\label{Fig4}
\end{figure}

Figure~\ref{Fig4} shows the energy dependence of ${\it{S}}\sigma$ and $\kappa$$\sigma^2$ for $\Delta N_{p}$ 
for Au+Au collisions for two collision centralities (0-5\% and 70-80\%), corrected for $p$($\bar{p}$) reconstruction 
efficiency. 
The ${\it{S}}\sigma$ values normalized to the corresponding Skellam expectations are shown in
the bottom panel of Fig.~\ref{Fig4}.  
The Skellam expectations reflect a system of totally uncorrelated, statistically random particle production. 
The corresponding results from the $p$+$p$ collisions are also shown and found to be similar to peripheral Au+Au 
collisions for $\sqrt{s_{\mathrm {NN}}}$ = 62.4 and 200 GeV within the statistical errors.
For $\sqrt{s_{\mathrm {NN}}}$ below 39 GeV, differences are observed between the 0-5\% central Au+Au collisions and 
the peripheral collisions. The results are closer to unity for $\sqrt{s_{\mathrm {NN}}}$ = 7.7 GeV. 
Deviations of 0-5\% Au+Au data from Skellam expectations, 
($(\mid {\rm Data} - {\rm Skellam} \mid)/\sqrt{\rm {err_{stat}}^{2} + {\rm err_{sys}}^{2}}$) 
are found to be most significant for 19.6 GeV and 27 GeV, with values of 3.2 and 3.4
for $\kappa$$\sigma^2$, and 4.5 and 5.6 for ${\it{S}}\sigma$, respectively.
The deviations for 5-10\% Au+Au data are smaller for $\kappa$$\sigma^2$  with values of 2.0 and 0.6 
and are 5.0 and 5.4 for ${\it{S}}\sigma$, for 19.6 GeV and 27 GeV,  respectively.
A reasonable description of the measurements is obtained from the independent production approach.
The data also show deviations from the hadron resonance gas model~\cite{Karsch:2010ck,Garg:2013ata}
which predict $\kappa$$\sigma^2$ and  ${\it{S}}\sigma$/Skellam to be unity.
To understand the effects of baryon number conservation~\cite{Bzdak:2012an} and experimental acceptance,
UrQMD model calculations (a transport model which does not include a CP)~\cite{urqmd} for 0-5\% Au+Au collisions 
are shown in the middle 
and bottom panels of Fig.~\ref{Fig4}. The UrQMD model shows a monotonic decrease with decreasing beam energy~\cite{Luo:2013bmi}.

The current data provide the most relevant measurements over the widest range in $\mu_{B}$ (20 to 450 MeV) 
to date for the CP search, and for comparison with the 
baryon number susceptibilities computed from QCD  to understand the various features 
of the QCD phase structure~\cite{qcp1,latticesus,Gavai:2010zn}.
The deviations of ${\it{S}}$$\sigma$ and $\kappa$$\sigma^2$ below Skellam expectation are qualitatively
consistent with a QCD based model which includes a CP~\cite{Stephanov:2011zz}. However the UrQMD 
model which does not include a CP also shows deviations from the Skellam expectation.
Hence conclusions on the existence of CP can be made only after comparison to QCD calculations with 
CP behavior which include the dynamics associated with heavy-ion collisions, such as finite 
correlation length and freeze-out effects. 

In summary, measurements of the higher moments and their products (${\it{S}}$$\sigma$ and $\kappa$$\sigma^2$) 
of the net-proton distributions 
at midrapidity ($|y|$$<$ 0.5) within 0.4 $<$ $p_{\mathrm T}$ $<$ 0.8 GeV/$c$ 
in Au+Au collisions over a wide range of  $\sqrt{s_{\mathrm {NN}}}$ and $\mu_{\mathrm B}$ 
have been presented to search for a possible CP and signals of a phase transition
in the collisions. 
These observables show a centrality and energy dependence, which
are neither reproduced by non-CP transport model calculations, nor by a hadron resonance gas model. 
For $\sqrt{s_{\mathrm {NN}}}$ $>$ 39 GeV,  ${\it{S}}$$\sigma$ and $\kappa$$\sigma^2$  
values are similar for central, peripheral Au+Au collisions and $p$+$p$ collisions. 
Deviations for both $\kappa$$\sigma^2$ and ${\it{S}}$$\sigma$ from HRG and Skellam expectations are 
observed for $\sqrt{s_{\mathrm {NN}}}$  $\le$ 27 GeV. The measurements are reasonably described by 
assuming independent production of $N_{p}$ and $N_{{\bar{p}}}$, 
indicating that there are no apparent correlations between the protons and anti-protons
for the observable presented. However at the lower beam energies, the net-proton measurements are dominated
by the shape of the proton distributions only. The data presented here also provides information 
to extract freeze-out conditions in heavy-ion collisions using QCD based approaches~\cite{Bazavov:2012vg,Borsanyi:2013hza}.

\indent 
We thank M. Asakawa, R. Gavai, S. Gupta, F. Karsch, K. Rajagopal, K. Redlich and M. A. Stephanov for discussions related to this work.
We thank the RHIC Operations Group and RCF at BNL, and the NERSC Center 
at LBNL, the KISTI Center in Korea  and the Open Science Grid consortium 
for providing resources and support. This work was supported in part by the Offices of NP 
and HEP within the U.S. DOE Office of Science, the U.S. NSF, CNRS/IN2P3, 
FAPESP CNPq of Brazil, Ministry of Ed. and Sci. of the Russian Federation, 
NNSFC, CAS, MoST, and MoE of China, the Korean Research Foundation, GA and MSMT of the 
Czech Republic, FIAS of Germany, DAE, DST, and CSIR of the 
Government of India, National Science Centre of Poland, National Research Foundation (NRF-2012004024),
Ministry of Sci., Ed. and Sports of the Rep. of Croatia, and RosAtom of Russia.
Finally, we gratefully acknowledge a sponsored research grant for the 2006 run period from 
Renaissance Technologies Corporation.


\begin{thebibliography}{9}
\bibitem{eff} $ C_{2}^{X-Y} = ({C_{2}^{x-y} + (\varepsilon  - 1)( \langle x \rangle  +  \langle y \rangle )})/{{\varepsilon ^2}}$;
$C_{3}^{X - Y} = (C_{3}^{x-y} + 3(\varepsilon  - 1)(C_{2}^{x} - C_{2}^{y}) + (\varepsilon  - 1)(\varepsilon  - 2)( \langle x \rangle  -  \langle y \rangle ))/{\varepsilon ^3}$ 
and 
${C_4}(X - Y) = ({C_4}(x - y) - 2(\varepsilon  - 1){C_3}(x + y) + 8(\varepsilon  - 1)({C_3}(x) + {C_3}(y)) + (5 - \varepsilon )(\varepsilon  - 1){C_2}(x + y) + 8(\varepsilon  - 2)(\varepsilon  - 1)({C_2}(x) + {C_2}(y)) + ({\varepsilon ^2} - 6\varepsilon  + 6)(\varepsilon  - 1)( \langle x \rangle  +  \langle y \rangle ))/{\varepsilon ^4}$.
Where ($X,Y$) and ($x,y$) are the 
numbers of ($p$,$\bar{p}$) produced and measured respectively. $\varepsilon$ is the $p(\bar{p})$ reconstruction efficiency.
\end{thebibliography}
\end{document}